\documentclass[final,3p,times]{elsarticle}

\newcommand{\vek}[1]{\mathbf{#1}} 

\newcommand{\nn}[0]{\nonumber}

\newcommand{\mat}[1]{\underline{\underline{\mathbf{#1}}}}

\newcommand{\bfnabla}[0]{\boldsymbol{\nabla}}

\biboptions{comma,square,numbers,sort}
\journal{Journal of Computational Physics}

\usepackage{graphicx}
\usepackage{dcolumn}
\usepackage{amsbsy}
\usepackage{multirow}
\usepackage[abs]{overpic}
\usepackage{natbib}
\usepackage[T1]{fontenc}

\begin{document}

\begin{frontmatter}

\title{Simulating Viscous Fingering with a Timespace Method and Anisotropic Mesh Adaptation}
\author{Kristian Ejlebjerg Jensen\corref{cor1}}
\ead{kristianejlebjerg@gmail.com}
\address{Department of Earth Science and Engineering, Imperial College London, SW7 2AZ London, United Kingdom}
\cortext[cor1]{Corresponding author}

\begin{abstract}
We report findings related to a two dimensional viscous fingering problem solved with a timespace method and anisotropic elements. Timespace methods have attracted interest for solution of time dependent partial differential equations due to the implications of parallelism in the temporal dimension, but there are also attractive features in the context of anisotropic mesh adaptation; not only are heuristics and interpolation errors avoided, but slanted elements in timespace also correspond to long and accurate timesteps, i.e. the anisotropy in timespace can be exploited. We show that our timespace method is restricted by a minimum timestep size, which is due to the growth of numerical perturbations. The lower bound on the timestep is, however, quite high, which is indicative that the number of timesteps can be reduced with several orders of magnitude for practical applications.
\end{abstract}%

\end{frontmatter}

\section{Introduction}
Many engineering problems are governed by partial differential equations which can only be solved numerically. The prospect of parallelising such simulations in the time dimension have attracted attention by the scientific community \cite{maday2002parareal,bal2002parareal,gosselet2006non,farhat2003time}, and the parareal algorithm \cite{lions2001resolution} is popular and widely studied. The parallelism in the time dimension can be achieved by leaving the distinction between time and space, in what is normally referred to as timespace methods. The technique is often used for solving hyperbolic problems with dicontinuous Galerkin (DG) formulations \cite{thite2008efficient}, as demonstrated for the full four dimensional case in the context of rotor flows \cite{van2008adaptive}. For that application all the complications of a time varying geometry disappear, when switching to a timespace method. This advantage was also exploited for dynamic stall prediction as simulated with the Navier-Stokes equations \cite{klaij2006space}. That work utilised spacetime adaptation and, as it is often the case, a compromise was made in the form of timeslabs, which allows for a tunable problem size \cite{van2002space}. 

Anisotropic mesh adaptation is an established technique for ensuring computational efficiency in the context of multiscale problems \cite{loseille2010fully,habashi2000anisotropic,loseille2011continuous,pain2001tetrahedral}. There exists many heuristics methods for applying the technique to transient problems \cite{pain2001tetrahedral,alauzet20073d,li20053d}, but the only way to exploit the anisotropy in the physics as seen in timespace, is by using a timespace method. This is also a conceptually simple way to avoid the interpolation errors and heuristics associated with combining conventional timestepping algorithms with mesh adaptation. The benefits of anisotropic mesh adaptation generally increases with dimensionality due to the fact that elements can be stretched in more dimensions, and one could thus expect significant benefits from the use of a four dimensional anisotropic mesh adaptation algorithm, but (to our knowledge) anisotropic mesh adaptation has only been demonstrated for two and three dimensions. We are not aware of any studies involving the use of three dimensional anisotropic meshes for solving a timespace problem.

In this work, we focus on a miscible viscous fingering problem, which is transient and has two spatial dimensions. The problem is known to exhibit chaotic behaviour \cite{pramanik2012miscible}, which causes some issues for the combination of a timespace method and anisotropic mesh adaptation. We attribute this to the fact that the numerical noise can grow in time. This could mean that a non-linear solver is unable to converge to the error level that one would expect for the computational resources utilised. We suggest to use the method of timeslabs, where the slab thickness is a kind of timestep size. The method is used to investigate the relation between the timeslab thickness, the computational resources used and growth of numerical errors. 

\section{Anisotropic Mesh Adaptation} \label{sec:adapt}
\begin{figure}[!htb]
\centering
\includegraphics[width=0.35\textwidth]{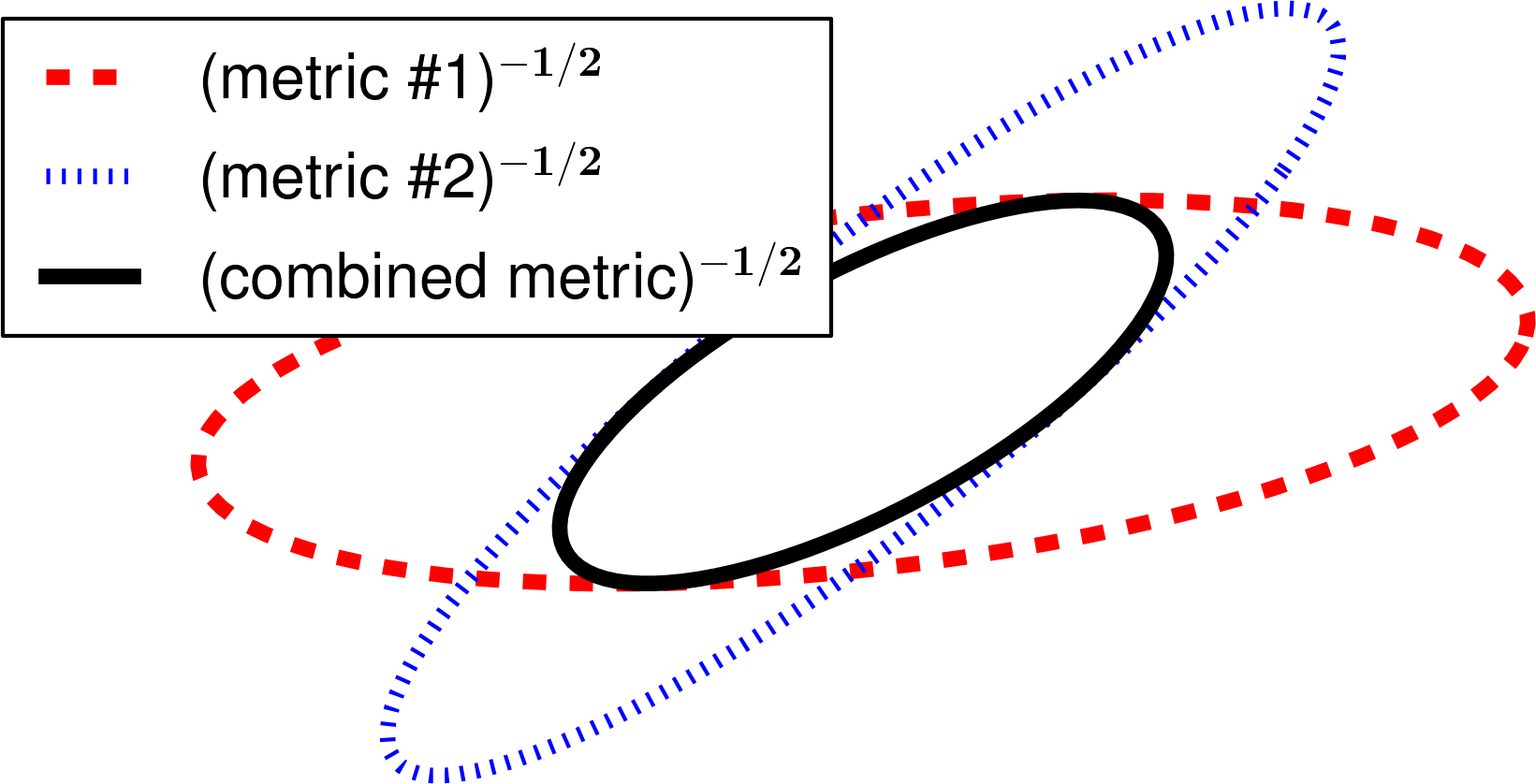}
\caption{The inner ellipse method is illustrated in the case of intersection, but it is common to see one ellipse entirely within the other, such that anisotropy is preserved.}\label{fig:ellipse}
\end{figure}

Mesh adaptation is the art of choosing the discretisation that minimises the error level for a given computational cost. It is important to distinguish between p- and h-adaptation, which relate to the discretisation order and length scale, respectively. Anisotropic mesh adaptation is the most general type of h-adaptation, because it not only considers the local element size, but also the element orientation. The optimal size and orientation can be expressed in terms of a metric tensor field \cite{loseille2011continuous}, $\mat{\mathcal{M}}$. This has units of inverse square length, and it can be used to map elements to metric space, where the optimal element has unit edge lengths. The metric that minimises the interpolation error \cite{chen2007optimal} of a variable with Hessian, $\mat{H}$, is 
\begin{eqnarray} 
\mat{\mathcal{M}} = \frac{1}{\sigma}\left(\mathrm{det}[\mat{\mathrm{abs}}(\mat{H})]\right)^{-\frac{1}{2q+d}} \mat{\mathrm{abs}}(\mat{H}) \label{eqn:M} ,  
\end{eqnarray}
where $d$ is the number of dimensions, $q$ is the error norm to be minimised, $\mathrm{det}$ is the determinant, $\sigma$ is a scaling factor and $\mat{\mathrm{abs}}$ takes the absolute value in the principal frame. The metrics of several variables can be combined with the inner ellipsoid method \cite{pain2001tetrahedral}, illustrated for two dimensions in figure \ref{fig:ellipse}. 

We use 1st order polynomials for all fields, so we compute the Hessian by first performing a Galerkin projection of the gradient onto 1st order polynomials and then repeating the process to get a nodal Hessian. The nodal metric can then be calculated explicitly using equation (\ref{eqn:M}).

\begin{figure}[!htb]
\centering
\includegraphics[width=\textwidth]{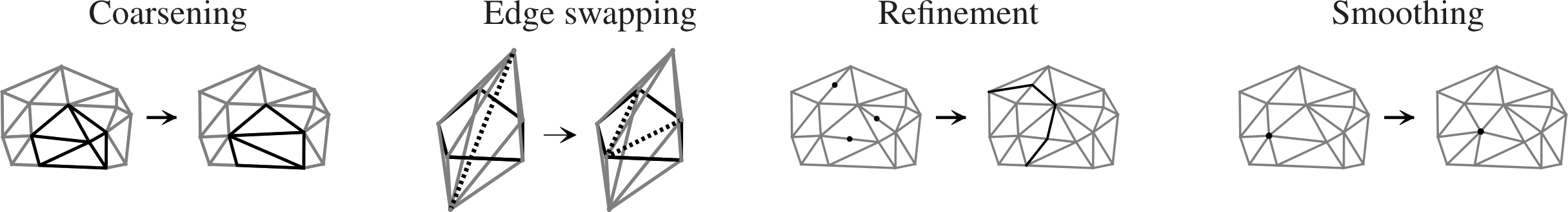}
\caption{Four local mesh modifications are illustrated: Coarsening, refinement, swapping  and smoothing. Only the edge swapping operation is illustrated in three dimensions, as it is trivial to generalise the other operations to three dimensions. Coarsening is the only operation that is allowed to reduce the worst local element quality and therefore edges marked for refinement are not guaranteed to be split. We do not use face to edge operations also known as 2-to-3 operations.}\label{fig:meshmod}
\end{figure}

We use the technique of local mesh modifications to adapt the mesh to the metric. There are four operation types: Coarsening, swapping, refinement and smoothing, as shown in figure \ref{fig:meshmod}. Coarsening removes edges that are short in metric space without regard to mesh quality, but the other modifications only allow the worst local element quality to go up. The mesh quality is quantified by means of the Vassilevski functional \cite{vasilevski2005error}.

We use an Octave/MATLAB implementation, which is suboptimal in terms of performance, especially compared to a similar C++ implementation \cite{rokos2013thread}. 

\section{Setup}
We consider a simple two-phase miscible problem \cite{pramanik2012miscible}, where the viscosity, $\eta$, depends exponentially on the saturation, $\phi$, with $\phi = 0$ and $\phi=1$ corresponding to the low and high viscosity fluid, respectively.
\begin{eqnarray}
\eta(\phi) &=& \eta_0e^{\xi\phi}, \label{eqn:visc}
\end{eqnarray}
where $\xi$ determines the viscosity ratio. The velocity, $\vek{v}$, is given by the Darcy equation,
\begin{eqnarray}
\vek{v} &=& -\frac{\kappa}{\eta(\phi)} \bfnabla p, \label{eqn:darcy}
\end{eqnarray}
where $p$ is the pressure and $\kappa$ is the permeability. Mass conservation leads to a Poisson equation for the pressure,
\begin{eqnarray}
0 = \bfnabla \cdot (\rho \vek{v}) &=& \bfnabla \cdot \left(\rho \frac{\kappa}{\eta(\phi)}\bfnabla p \right) \nn \\
&=& \bfnabla \cdot \left(\frac{\kappa}{\eta(\phi)}\bfnabla p \right)  ,\label{eqn:p}
\end{eqnarray}
where $\rho$ is the density. This drops out of the equations, because we  choose to study the case of equal densities for the two fluids. The saturation is convected with the velocity given in equation (\ref{eqn:darcy}),
\begin{eqnarray}
0 &=& \frac{\partial \phi}{\partial t} + \vek{v} \cdot \bfnabla \phi \nn \\
&=& \frac{\partial \phi}{\partial t} - \frac{\kappa}{\eta(\phi)} \bfnabla p \cdot \bfnabla \phi \label{eqn:phi}
\end{eqnarray}
The equation is treated as a convective equation with a three dimensional velocity vector, and it is stabilised with streamline upwind Petrov/Galerkin diffusion. The local characteristic length scale for the mesh is calculated by computing the Steiner ellipsoid for every element and projecting this along the velocity direction.
 
\begin{figure}[!htb]
\centering
\includegraphics[width=0.4\textwidth]{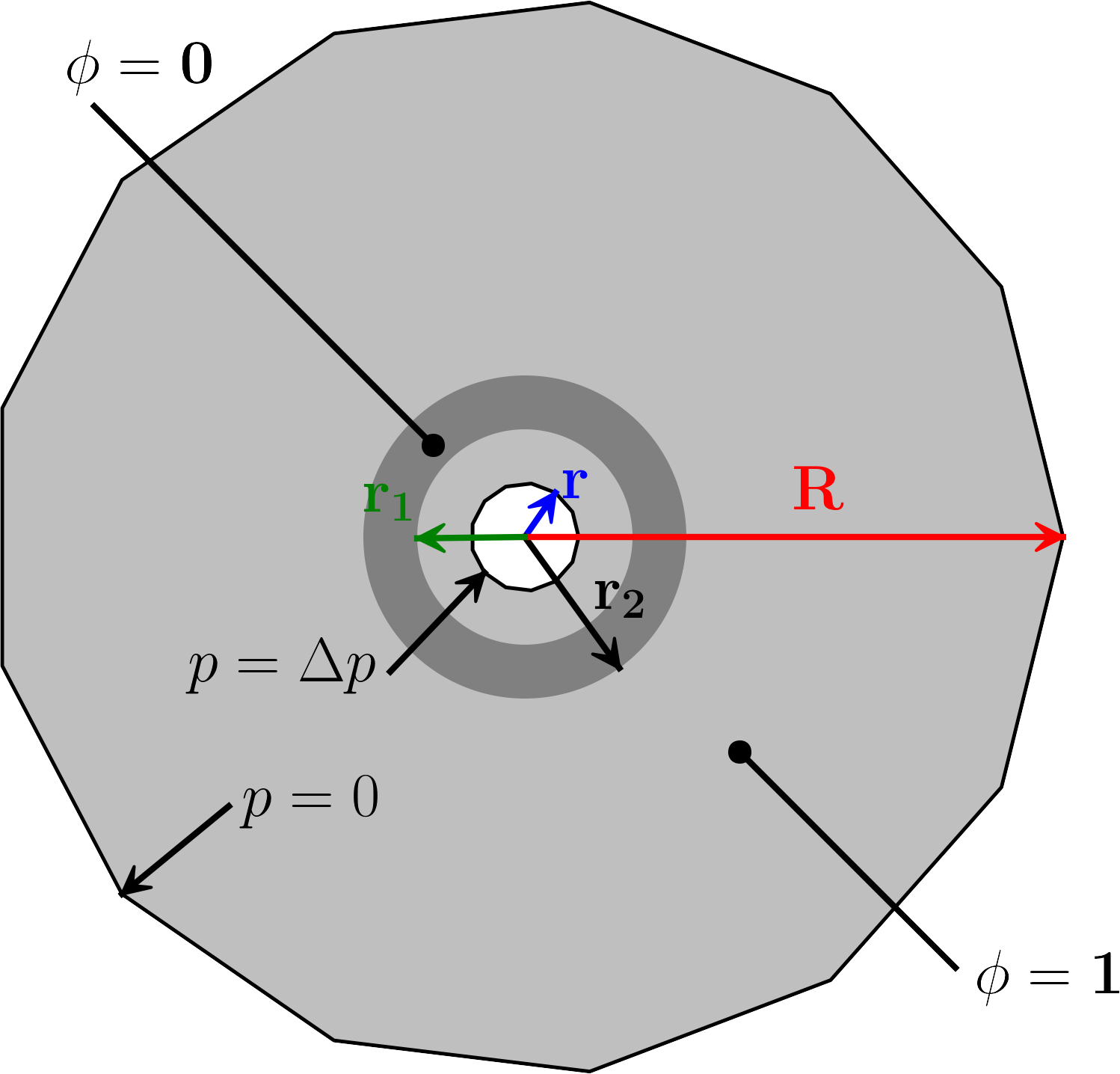}
\caption{The initial condition for the viscous fingering is sketched  with the dark grey fluid pushing out on the light gray fluid, which has a higher viscosity. The initial saturation is unperturbed and it is thus primarily the length scale of the initial mesh that triggers the instability. Zero pressure is prescribed at the outer boundary, while $p=\Delta p$ and $\phi=1$ are prescribed at the inlet.}\label{fig:setup}
\end{figure}

The governing equations (\ref{eqn:p}) and (\ref{eqn:phi}) are solved with the boundary conditions illustrated in figure \ref{fig:setup}. The use of a 13-sided polygon to approximate a circle avoids extrapolation, when comparing solution fields on different meshes. The perfect circular initial condition for the saturation makes the solution undefined, which causes the growth of numerical perturbations to become very obvious.  All equations are solved with the finite element method using FEniCS \cite{LoggMardalEtAl2012a}, an open source finite element package. Both pressure and saturation is discretised with first order polynomials. A direct solver (LU) is used for equations (\ref{eqn:p}) and (\ref{eqn:phi}), while an iterative solver (CG+ILU) is used for Galerkin projections. All computations are single threaded. 

We non-dimensionalise the equations with $\Delta p$ as characteristic pressure and $R$ as characteristic length scale,
\begin{eqnarray}
\vek{x} &=& R \tilde{\vek{x}}, \quad t = \frac{R^2\eta_0}{\tilde{v}_\mathrm{char} \kappa \Delta p} \tilde{t}, \quad p=\Delta p \tilde{p}, \nn \\
r &=& 0.1R, \quad r_1 = 0.2R, \quad r_2 = 0.3R, \quad \xi = 2, \quad \tilde{t}_\mathrm{final} = 1 \nn
\end{eqnarray}
where $\tilde{\vek{x}}$, $\tilde{t}$, $\tilde{p}$ are dimensionless space, time and pressure, respectively. $\tilde{t}_\mathrm{final}$ is the simulation time. $\tilde{v}_\mathrm{char}$ is a numerical parameter, which determines the relative resolution of space and time. We fix it at unity. Another numerical parameters is the norm of the interpolation error to be minimised, and we go with the popular choice of $q=2$ for this. 

The saturation at the initial time is imposed by means of a Dirichlet boundary condition. We initialize the non-linear solver with $\phi=0$ and $\tilde{p}=\tilde{p}_\mathrm{init}$, where
\begin{eqnarray}
\tilde{p}_\mathrm{init} &=& \frac{\tilde{r} - 0.1}{0.9}, \nn
\end{eqnarray}

\begin{figure}[!htb]
\centering
\includegraphics[height=0.45\textwidth]{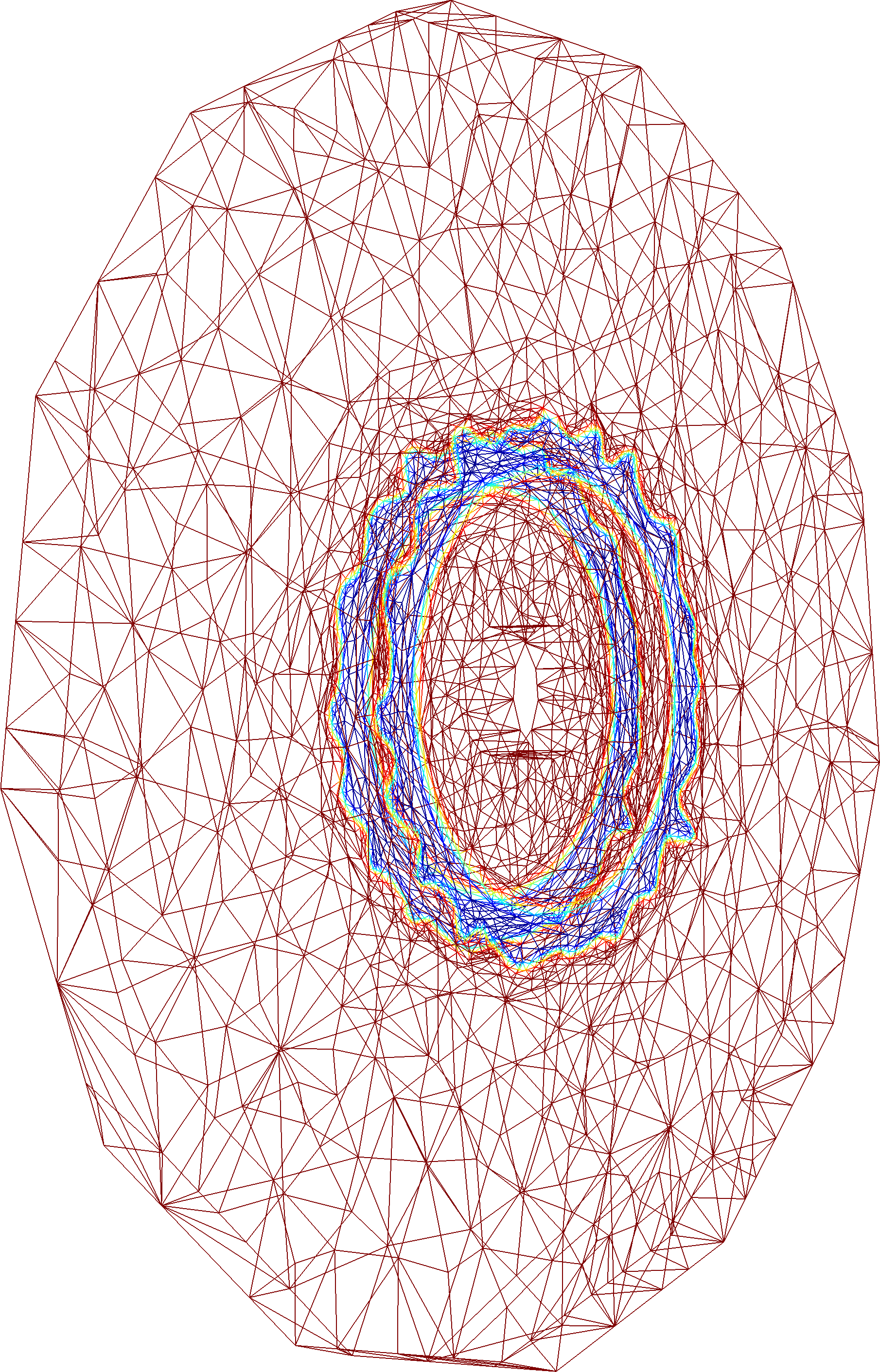} \hspace{0.06\textwidth}
\includegraphics[height=0.45\textwidth]{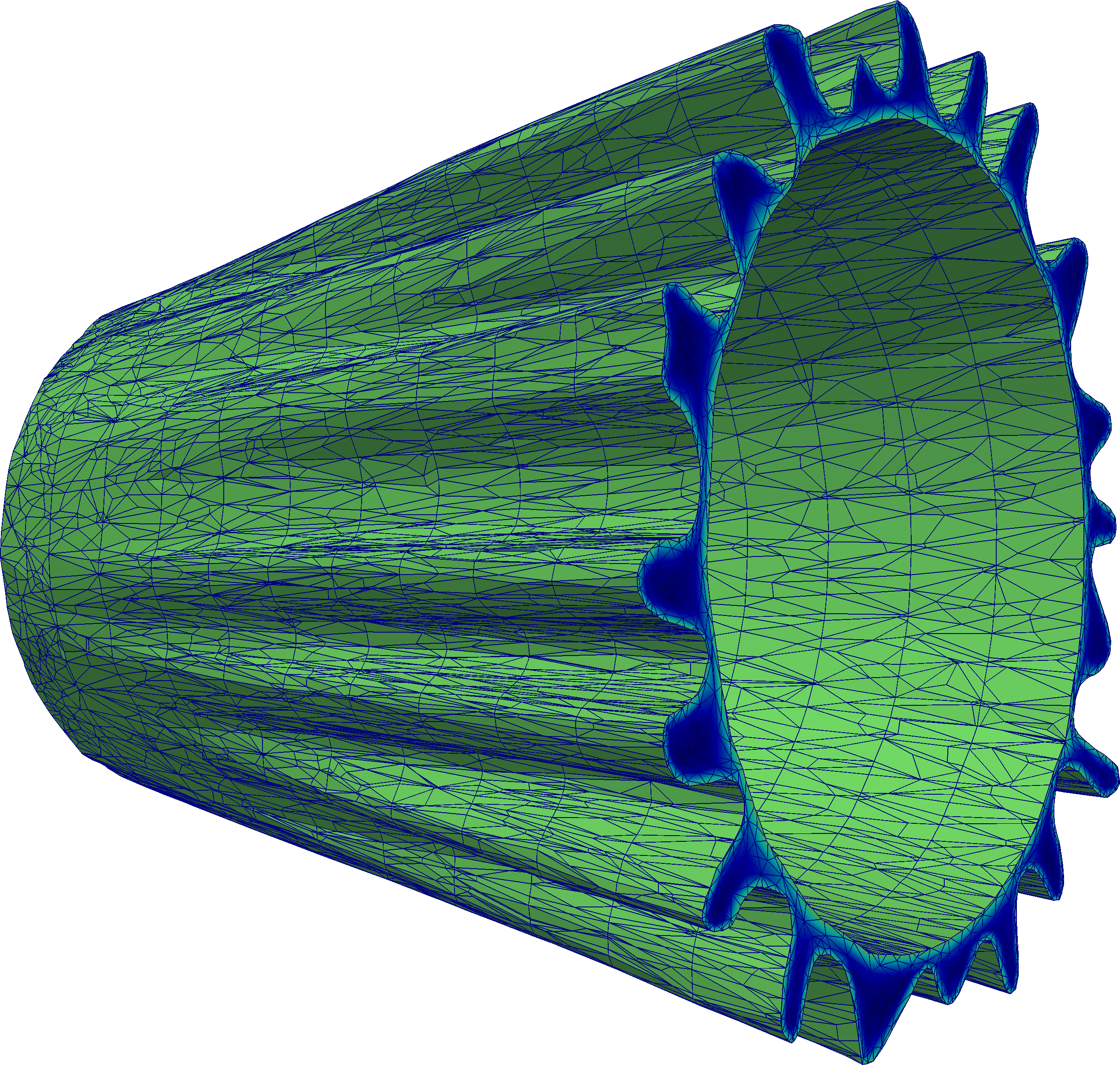} 
\caption{A simulation with $\sigma=0.02$ and $\Delta t=0.1$ is plotted in terms of a wireframe of the final iteration with slab \#5 (left) and all slabs in terms of the $\phi=0.5$ isosurface (right). The time dimension is from left to right pointing out of the plane. Large elements clearly show up away from the interface, but even at the interface the anisotropy in timespace can be exploited such that very few elements are needed. Note that the mesh is fully tetrahedral; the quadrilaterals on the isosurface representation appear, when cutting tetrahedrons with two nodes on both sides of the threshold.}
\label{fig:funky}
\end{figure}

$\tilde{r}$ being the radial coordinate. We use a segregated fixed point method to deal with the non-linearity of the equations. The timespace problem is solved in slabs with a thickness of $\Delta t$ (omitting the tilde), i.e.
\begin{itemize}
\item[\#0] Initialise mesh, and set $\phi=0$, $\tilde{p}=\tilde{p}_\mathrm{init}$.
\item[\#1] Solve equation (\ref{eqn:p}) for the pressure with fixed saturation.
\item[\#2] Solve equation (\ref{eqn:phi}) for the saturation by taking the viscosity to be constant using the previous saturation.
\item[\#3] If it is the first iteration at this timeslab, go to \#1. Otherwise, we calculate metric fields associated with the pressure and the saturation. These are combined with the inner ellipsoid method (see figure \ref{fig:ellipse}), the mesh is updated and the fields are interpolated onto the new mesh. If it is not the first timeslab, the mesh is fixed at the boundary matching up to the previous timeslab, 
\item[\#4] If 20 iterations have been carried out with the current timeslab, we proceed to the next one by mirroring the mesh in a plane normal to the time direction, and set $\phi=0$, $\tilde{p}=\tilde{p}_\mathrm{init}$. We then go to \#1.
\end{itemize}
The mesh is changed in every iteration, and therefore the non-linear solver cannot converge to machine precision. A typical timeslab is shown in figure \ref{fig:funky} together with the result of a full timespace simulation. Both figures illustrate the value of full timespace anisotropy.

\section{Results}
\begin{figure}[!htb]
\centering
\includegraphics[width=0.3\textwidth]{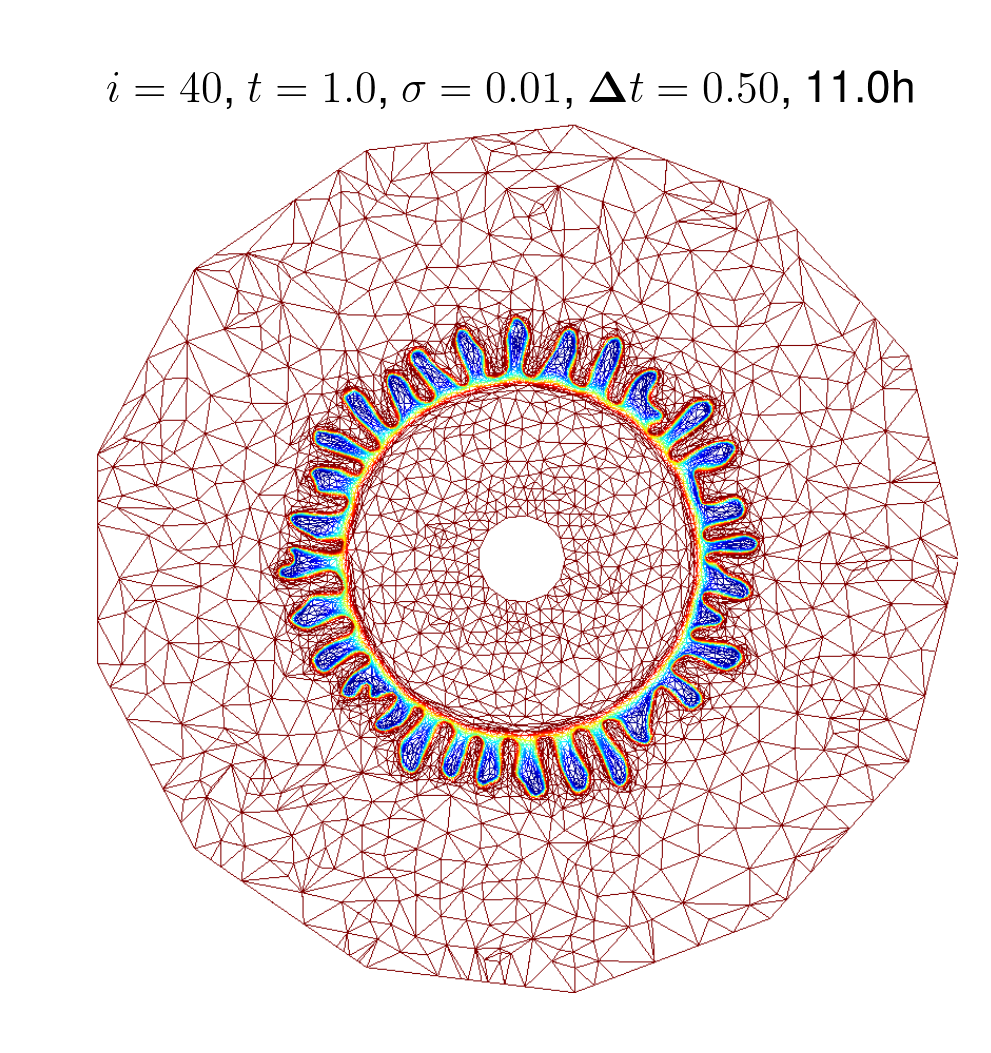} \hspace{0.03\textwidth}
\includegraphics[width=0.3\textwidth]{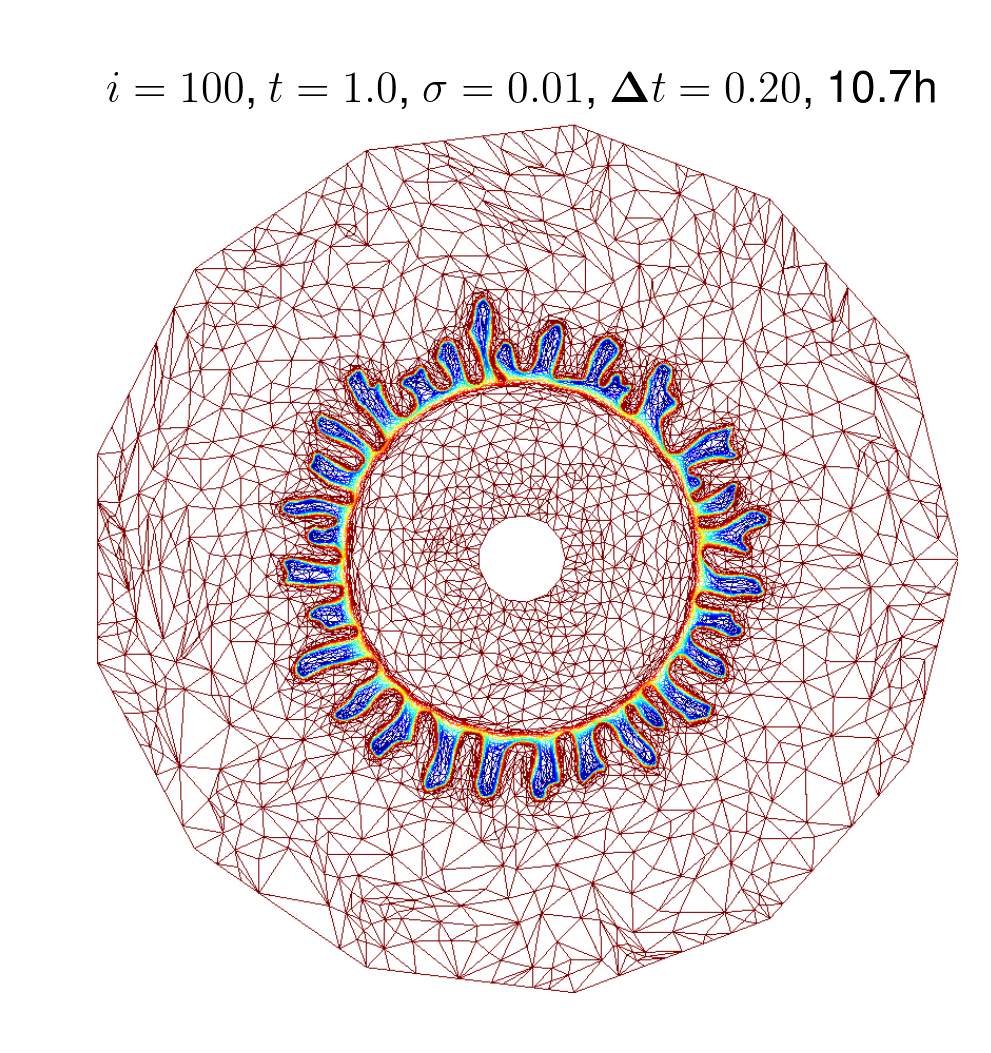}\\
\includegraphics[width=0.3\textwidth]{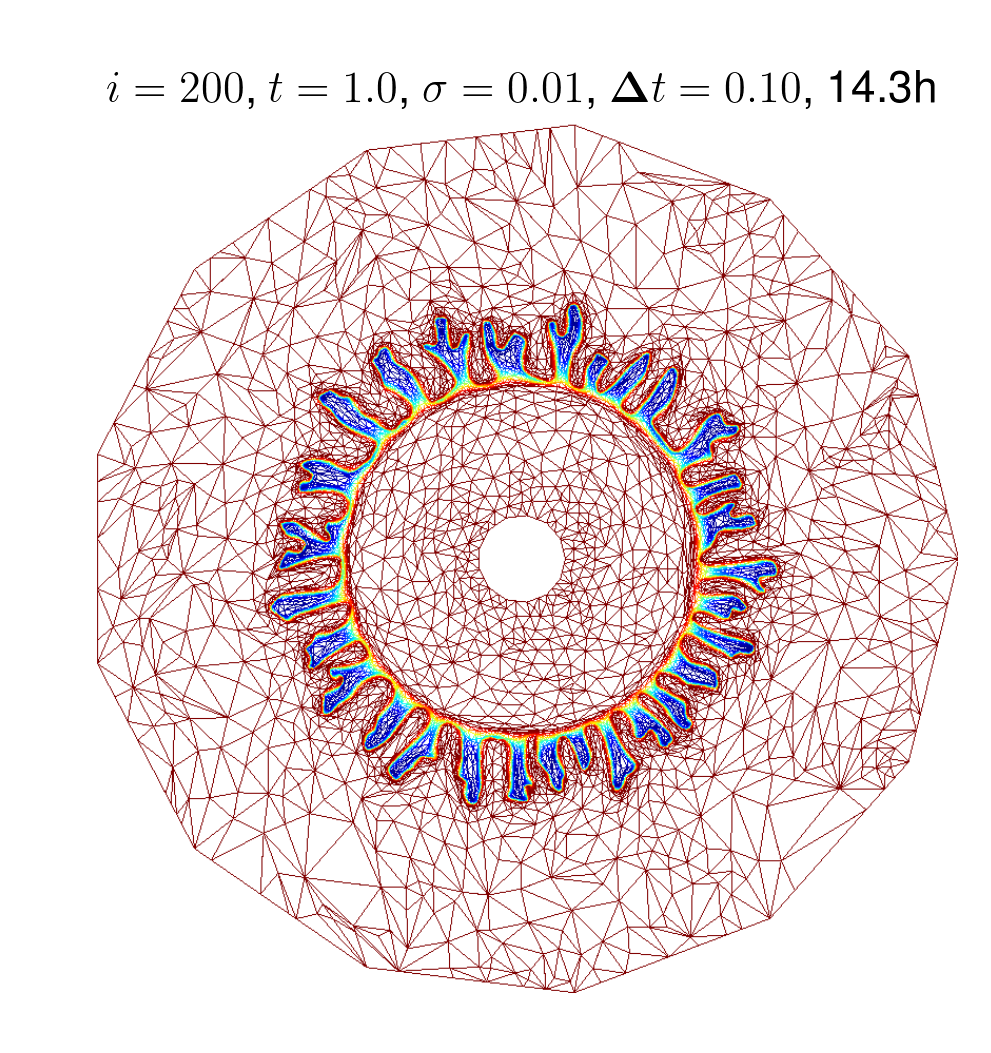} \hspace{0.03\textwidth}
\includegraphics[width=0.3\textwidth]{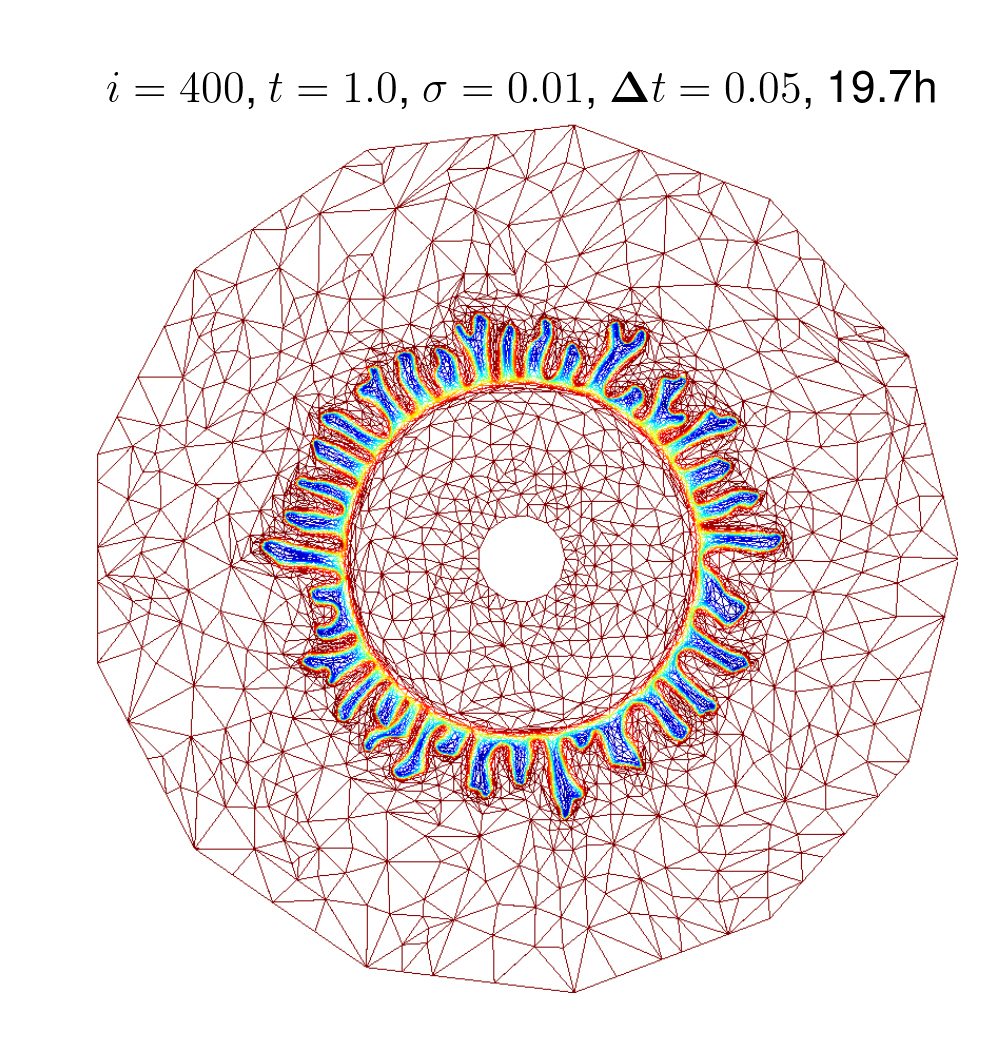}
\caption{The saturation at the final time is plotted for four simulations with $\sigma=0.01$ and $\Delta t$= 0.05, 0.1, 0.2 and 0.05. The shape of the fingers vary, but the number and length seems independent of $\Delta t$.}
\label{fig:dt}
\end{figure}

\begin{figure}[!htb]
\centering
\includegraphics[width=0.3\textwidth]{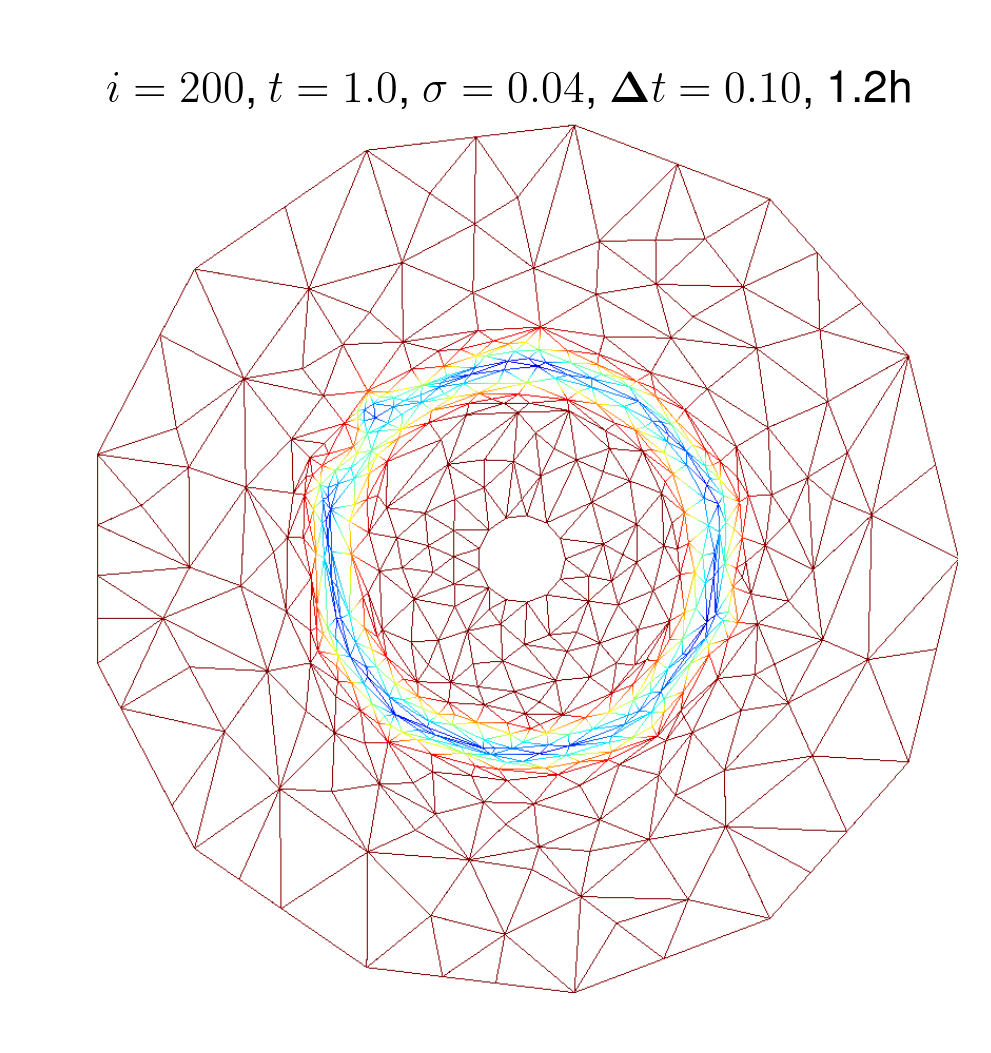} \hspace{0.03\textwidth}
\includegraphics[width=0.3\textwidth]{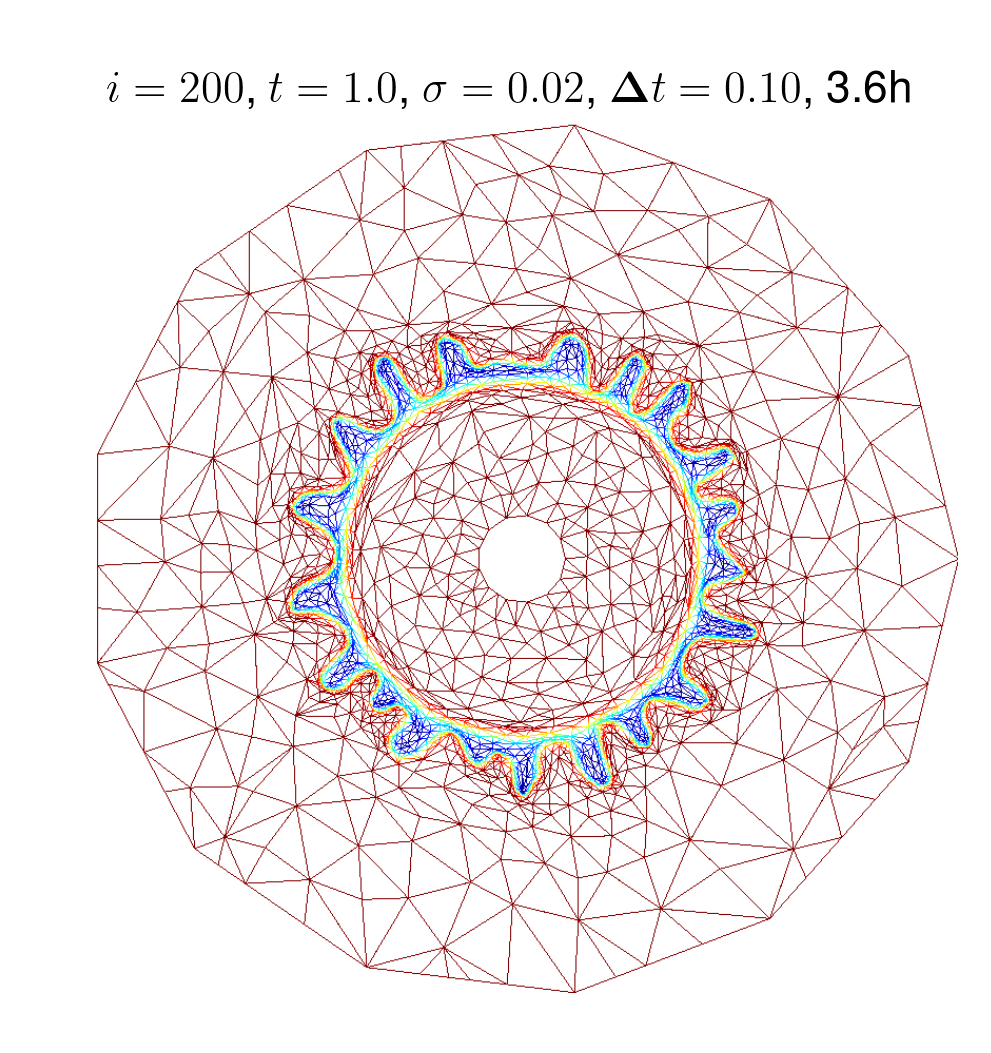} \hspace{0.03\textwidth}
\includegraphics[width=0.3\textwidth]{U_R_2_eta_1_Lz_1_pp1_p2_v1_sin0_dt10_oct4}
\caption{The saturation at the final time is plotted for three simulations with $\Delta t=0.1$ and $\sigma$ = 0.04, 0.02 and 0.01. Both the shape and number of fingers seem highly dependent on $\sigma$.}
\label{fig:eta}
\end{figure}

We show result for timeslab thicknesses $\Delta t$ = 0.05, 0.1, 0.2 and 0.5 with $\sigma=0.01$ as well as for mesh tolerances $\sigma$ = 0.04, 0.02 and 0.01 with $\Delta t=0.1$. The result at the end time is plotted in figures \ref{fig:dt} and \ref{fig:eta}, respectively. The final iteration number, time, $\Delta t$ value, $\sigma$ value and final total computational time is printed above each image. As one would expect, the shape and number of fingers clearly varies with $\sigma$, but not with $\Delta t$.

\begin{figure}[!htb]
\centering
\includegraphics[width=\textwidth]{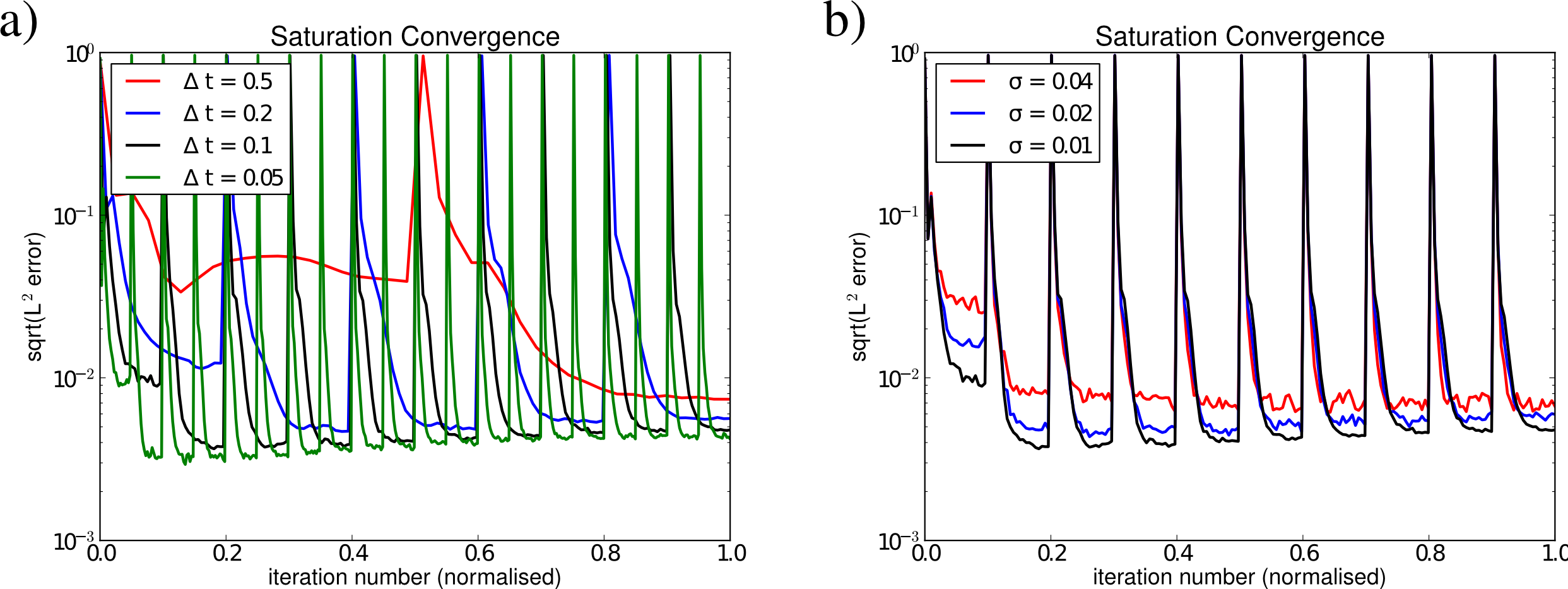}
\caption{The square root of the $L^2$ residual is plotted as a function of normalised iterations numbers. (a) shows varying $\Delta t$ at $\sigma=0.01$, while (b) shows varying $\sigma$ at $\Delta t=0.1$. The peaks occur whenever the timeslab is advanced in time.}\label{fig:conv}
\end{figure}

\begin{figure}[!htb]
\centering
\includegraphics[width=0.4\textwidth]{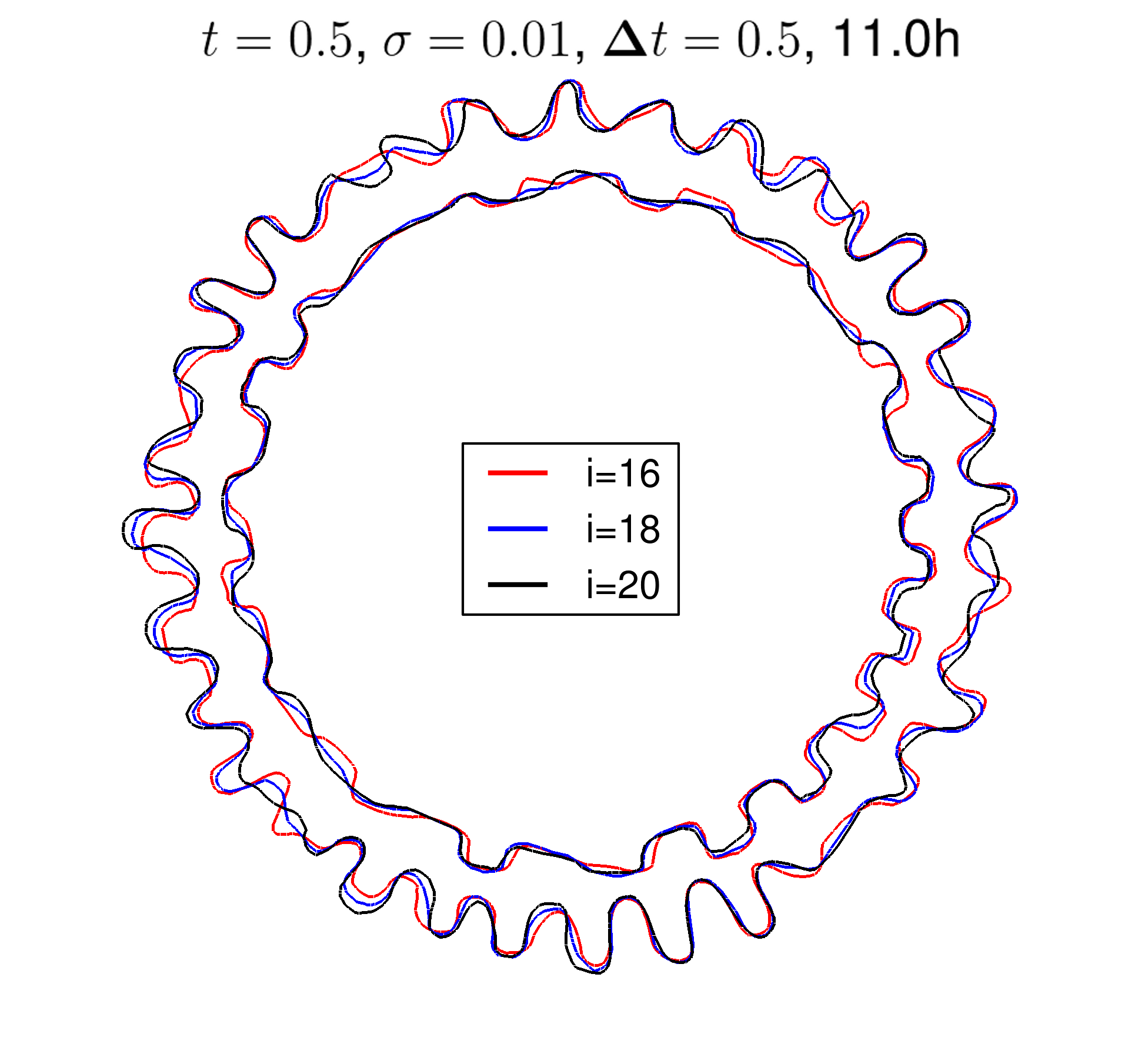} 
\caption{The $\phi=\tilde{t}=0.5$ contour is plotted for a simulation with $\Delta t=0.5$ and $\sigma=0.01$ for iteration numbers 16, 18 and 20. Note how different the fingers between iteration 16 and 20 are. We expect the mean radial position of the contours to be well defined, but not the actual fingers; they are determined by the mesh at the initial time, which changes at every iteration.}
\label{fig:nconv}
\end{figure}

We measure convergence in terms of the difference between saturation fields in consecutive iterations,
\begin{eqnarray}
\mathrm{L}^2_i = \frac{\int_\Omega (\phi_i -\phi_{i-1})^2 d\Omega}{\int_\Omega d\Omega} \nn .
\end{eqnarray}
This is plotted in figure \ref{fig:conv}(a-b). Figure \ref{fig:conv}(a) clearly shows that simulations become more accurate as $\Delta t$ is lowered, but the convergence is different when $\sigma$ is decreased, at least in the last half of the simulations: The error seems to scale with $\sigma$, but it scales down to a certain level with $\Delta t$. By comparing figure \ref{fig:conv}(a) and (b), we can see that this is the level of the overall discretisation error. We attribute this effect to the fact that perturbations due to the mesh grow with time due to the non-linear nature of the problem. The extreme case of this is illustrated in figure \ref{fig:nconv}, where snapshots of the $\phi=\tilde{t}=0.5$ contour is shown for three different iterations with $\Delta t=0.5$. There are significant differences due to the growth of mesh perturbations, particularly between iteration 16 and 20. This is to say, that the solution is not well defined and therefore it has an oscillatory component arising from the changing mesh. The error associated with this component increases with time, so it can be smaller than the overall discretisation error, if $\Delta t$ is sufficiently small. Focusing on this case with $\sigma=0.01$ and $\xi=2$, it looks like $\Delta t=0.1$ is a good value. 

\section{Conclusion}
The results show that it is possible to simulate chaotic phenomena with a timespace method. A CFL like condition still seems to exists, so the timeslab cannot be too thick. This has to do with the fact that the chaotic nature of the problem allows for a range of solutions and the difference between these solutions, should not exceed the overall discretisation error. The lower limit for the timeslab thickness, is high enough to indicate that the total number of timesteps can be reduced by orders of magnitude compared to a conventional method.

\section{Suggestions for Future Work}
It would be interesting to compare an established method against an optimised timespace implementation in the context of large scale simulations and anisotropic mesh adaptation. If practical problems are to be solved, a four dimensional anisotropic mesh adaptation algorithm will ultimately be needed. 

\section{Acknowledgement}
This work is supported by the Villum Foundation.

\section{References}

\begin{thebibliography}{20}
\providecommand{\natexlab}[1]{#1}
\providecommand{\url}[1]{\texttt{#1}}
\expandafter\ifx\csname urlstyle\endcsname\relax
  \providecommand{\doi}[1]{doi: #1}\else
  \providecommand{\doi}{doi: \begingroup \urlstyle{rm}\Url}\fi

\bibitem[Maday and Turinici(2002)]{maday2002parareal}
Yvon Maday and Gabriel Turinici.
\newblock A parareal in time procedure for the control of partial differential
  equations.
\newblock \emph{Comptes Rendus Mathematique}, 335\penalty0 (4):\penalty0
  387--392, 2002.

\bibitem[Bal and Maday(2002)]{bal2002parareal}
Guillaume Bal and Yvon Maday.
\newblock A "parareal" time discretization for non-linear pde's with
  application to the pricing of an american put.
\newblock In \emph{Recent developments in domain decomposition methods}, pages
  189--202. Springer, 2002.

\bibitem[Gosselet and Rey(2006)]{gosselet2006non}
Pierre Gosselet and Christian Rey.
\newblock Non-overlapping domain decomposition methods in structural mechanics.
\newblock \emph{Archives of computational methods in engineering}, 13\penalty0
  (4):\penalty0 515--572, 2006.

\bibitem[Farhat and Chandesris(2003)]{farhat2003time}
Charbel Farhat and Marion Chandesris.
\newblock Time-decomposed parallel time-integrators: theory and feasibility
  studies for fluid, structure, and fluid--structure applications.
\newblock \emph{International Journal for Numerical Methods in Engineering},
  58\penalty0 (9):\penalty0 1397--1434, 2003.

\bibitem[Lions et~al.(2001)Lions, Maday, and Turinici]{lions2001resolution}
Jacques-Louis Lions, Yvon Maday, and Gabriel Turinici.
\newblock R{\'e}solution d'edp par un sch{\'e}ma en temps
  {\guillemotleft}parar{\'e}el{\guillemotright}.
\newblock \emph{Comptes Rendus de l'Acad{\'e}mie des Sciences-Series
  I-Mathematics}, 332\penalty0 (7):\penalty0 661--668, 2001.

\bibitem[Thite(2008)]{thite2008efficient}
Shripad Thite.
\newblock Efficient spacetime meshing with nonlocal cone constraints.
\newblock \emph{arXiv preprint arXiv:0804.0946}, 2008.

\bibitem[van~der Ven(2008)]{van2008adaptive}
Harmen van~der Ven.
\newblock An adaptive multitime multigrid algorithm for time-periodic flow
  simulations.
\newblock \emph{Journal of Computational Physics}, 227\penalty0 (10):\penalty0
  5286--5303, 2008.

\bibitem[Klaij et~al.(2006)Klaij, van~der Vegt, and van~der
  Ven]{klaij2006space}
Christiaan~M Klaij, Jaap~JW van~der Vegt, and Harmen van~der Ven.
\newblock Space--time discontinuous galerkin method for the compressible
  navier--stokes equations.
\newblock \emph{Journal of Computational Physics}, 217\penalty0 (2):\penalty0
  589--611, 2006.

\bibitem[Van~der Vegt and Van~der Ven(2002)]{van2002space}
JJW Van~der Vegt and H~Van~der Ven.
\newblock Space--time discontinuous galerkin finite element method with dynamic
  grid motion for inviscid compressible flows: I. general formulation.
\newblock \emph{Journal of Computational Physics}, 182\penalty0 (2):\penalty0
  546--585, 2002.

\bibitem[Loseille et~al.(2010)Loseille, Dervieux, and
  Alauzet]{loseille2010fully}
Adrien Loseille, Alain Dervieux, and Fr{\'e}d{\'e}ric Alauzet.
\newblock Fully anisotropic goal-oriented mesh adaptation for 3d steady euler
  equations.
\newblock \emph{Journal of computational physics}, 229\penalty0 (8):\penalty0
  2866--2897, 2010.

\bibitem[Habashi et~al.(2000)Habashi, Dompierre, Bourgault, Ait-Ali-Yahia,
  Fortin, and Vallet]{habashi2000anisotropic}
Wagdi~G Habashi, Julien Dompierre, Yves Bourgault, Djaffar Ait-Ali-Yahia,
  Michel Fortin, and Marie-Gabrielle Vallet.
\newblock Anisotropic mesh adaptation: Towards user-independent,
  mesh-independent and solver-independent cfd. part i: General principles.
\newblock \emph{International Journal for Numerical Methods in Fluids},
  32\penalty0 (6):\penalty0 725--744, 2000.

\bibitem[Loseille and Alauzet(2011)]{loseille2011continuous}
Adrien Loseille and Fr{\'e}d{\'e}ric Alauzet.
\newblock Continuous mesh framework part i: well-posed continuous interpolation
  error.
\newblock \emph{SIAM Journal on Numerical Analysis}, 49\penalty0 (1):\penalty0
  38--60, 2011.

\bibitem[Pain et~al.(2001)Pain, Umpleby, De~Oliveira, and
  Goddard]{pain2001tetrahedral}
CC~Pain, AP~Umpleby, CRE De~Oliveira, and AJH Goddard.
\newblock Tetrahedral mesh optimisation and adaptivity for steady-state and
  transient finite element calculations.
\newblock \emph{Computer Methods in Applied Mechanics and Engineering},
  190\penalty0 (29):\penalty0 3771--3796, 2001.

\bibitem[Alauzet et~al.(2007)Alauzet, Frey, George, and
  Mohammadi]{alauzet20073d}
Fr{\'e}d{\'e}ric Alauzet, Pascal~J Frey, Paul-Louis George, and Bijan
  Mohammadi.
\newblock 3d transient fixed point mesh adaptation for time-dependent problems:
  Application to cfd simulations.
\newblock \emph{Journal of Computational Physics}, 222\penalty0 (2):\penalty0
  592--623, 2007.

\bibitem[Li et~al.(2005)Li, Shephard, and Beall]{li20053d}
Xiangrong Li, Mark~S Shephard, and Mark~W Beall.
\newblock 3d anisotropic mesh adaptation by mesh modification.
\newblock \emph{Computer methods in applied mechanics and engineering},
  194\penalty0 (48):\penalty0 4915--4950, 2005.

\bibitem[Pramanik et~al.(2012)Pramanik, Kulukuru, and
  Mishra]{pramanik2012miscible}
Satyajit Pramanik, GL~Kulukuru, and Manoranjan Mishra.
\newblock Miscible viscous fingering: Application in chromatographic columns
  and aquifers.
\newblock In \emph{COMSOL conference, Bangalore}, 2012.

\bibitem[Chen et~al.(2007)Chen, Sun, and Xu]{chen2007optimal}
Long Chen, Pengtao Sun, and Jinchao Xu.
\newblock Optimal anisotropic meshes for minimizing interpolation errors in
  $l^p$-norm.
\newblock \emph{Mathematics of Computation}, 76\penalty0 (257):\penalty0
  179--204, 2007.

\bibitem[Vasilevski and Lipnikov(2005)]{vasilevski2005error}
Yu~V Vasilevski and KN~Lipnikov.
\newblock Error bounds for controllable adaptive algorithms based on a hessian
  recovery.
\newblock \emph{Computational Mathematics and Mathematical Physics},
  45\penalty0 (8):\penalty0 1374--1384, 2005.

\bibitem[Rokos et~al.(2013)Rokos, Gorman, Southern, and Kelly]{rokos2013thread}
Georgios Rokos, Gerard~J Gorman, James Southern, and Paul~HJ Kelly.
\newblock A thread-parallel algorithm for anisotropic mesh adaptation.
\newblock \emph{arXiv preprint arXiv:1308.2480}, 2013.

\bibitem[Logg et~al.(2012)Logg, Mardal, Wells, et~al.]{LoggMardalEtAl2012a}
Anders Logg, Kent-Andre Mardal, Garth~N. Wells, et~al.
\newblock \emph{Automated Solution of Differential Equations by the Finite
  Element Method}.
\newblock Springer, 2012.
\newblock ISBN 978-3-642-23098-1.
\newblock \doi{10.1007/978-3-642-23099-8}.

\end{thebibliography}
\providecommand{\noopsort}[1]{}\providecommand{\singleletter}[1]{#1}%

\end{document}